# Do you see what I see? Taking perspective of others using facial images

Yustinus Eko Soelistio
Information System Department
Universitas Multimedia Nusantara
Tangerang, Indonesia
yustinus.eko@umn.ac.id

*Abstract*—Albeit many HCI / emotion recognition studies use facial expressive images, few scrutinize the accuracies of the people (experimenters and participants) in perceiving the expressions representing the intended emotions. The misinterpretation of the expression will put bias in the data and introduce questions on the validity of the studies. The possibility of misinterpretation of the expressions will be the focus of the experiment conducted in this study. The experiment will evaluate the ability of people in taking the perspective of others in spite of their current emotions and gender, and whether the expressions can be universally perceived. This study find that it is relatively safe to use facial expressive images for research as long as the emotions are exclusively within the six basic emotions.

*Keywords—perspective taking, emotion perception, emotion recognition*

## I. Introduction

Expressive facial images/videos have been used in many research areas including in HCI, specifically in the field of emotion recognition (e.g. [1-4, 28-29]). Most of these studies report satisfactory results in their experiments based on their chosen data. However, although these studies are able to prove solid methods and evaluation results, small efforts are given in describing proves of the validity of the facial images. Even studies in the area of psychology that use facial images to elicit emotions (e.g. [5-8]), most do not put enough explanation on the effectiveness of the images.

The lack of explanations pose questions on the validity of the methods (from the HCI and emotion recognition studies) where the evaluation results of the methods are heavily depended on the assumption that the facial images are correctly referred to the targeted emotion label, which could not always be the case (cf. [9-13]). One basic question on using facial images for emotion experiments is whether the users (experimenters/participants) are able to correctly assign the intended emotions as depicted in the images. Inaccurate assignment would produce distorted data which ultimately introduce bias in the experiments.

This study intends to explore the possibility of assigning inaccurate emotion from the facial images in researches. The exploration focuses on three possible causes of inaccuracy that introduce the following three research questions (RQ): (1) do participants able to take the perspective of other? (2) do participants collectively appoint to the same emotions shown in the images? and (3) do men and women appoint to the same emotions depicted by the images?

This article reports the answers of the RQs and the experiment conducted in the study. By answering the RQs, this study contributes in two ways: (1) confirming the effectiveness of using facial images in eliciting emotions in HCI/emotion recognition studies, and (2) exploring possibilities that not all emotions are equally perceived with the same intensities.

This article is organized as follows. Section II explains the logical chronology of the RQs based on existing studies. Section III describes the experiment setup and the hypotheses derived from the RQs. Section IV and V present the results and the discussion of the experiment. Finally the conclusion of the study will be presented in section VI.

## II. Existing studies

Many studies assume that each of the expression in the facial images is exclusive for one type of emotion and can be recognize equally by all people. There are a number of studies that support this assumption. For example a study by [14] who suggest the link between perspective taking (cognitive empathy) and altruism that allow people to take the perspective of other people and to feel his/her feeling (empathy), and by [15] who find that people are able to feel what others are feeling regardless of the status of affiliation between them.

The two example studies ([14-15]) suggests that people are able to feel what the other are feeling and therefore understand the type of emotion the other people are experiencing. However, some more recent studies suggests that there are factors preceding the process of perspective taking so that the experience of taking perspective of others (i.e. knowing what others are feeling) is not independent, such as suggested by [16-19].

A study from [16] finds some valuing process antecedent the feeling of empathy (as an other-oriented emotion response perceived from others' welfare). The experiments show that people are having deeper feeling of other's wellness when they give higher values to the person/event of interest. This implicates a personal enhancing or lessening the intensity of empathy. Similar finding is also demonstrated by [17] who prove that culture has a considerable impact on the ability of people to take the perspective of other and self (cognitive empathy). In their finding, people grown in eastern culture are better in taking the perspective of others than those who grown in western culture. The differences are not only appear in the ability to perceive other's perspective, but also in the facial expressions [18] which makes the basis for RQ1: the effect of facial images should be differ between people based on their cultural background and their personal value on the image (i.e. the person portrayed in the image). The effect of personal value should also lessen the adherence of taking the perspective of other people's feeling (i.e. people have varied perspective on others' feeling), therefore makes the argument for RQ2. The argument for RQ3 is more straightforward where it bases solely on the finding of gender effect on empathy [19].

## III. METHODS

This study consists of an experiment divided into two steps: data acquisition and statistical analysis. The data acquisition step includes facial images preparation and survey. The data collected from the survey is analyzed based on statistical calculation to answer the RQs presented in Section 1.

### A. Data acquisition

The data is collected from 50 participants (25 participants for each gender). All participants are active students at Universitas Multimedia Nusantara between ages of 18-24 recruited voluntarily with a souvenir as incentive. They are presented by seven expressive facial images (Fig. 1) taken from [20] chosen randomly prior the experiment (without any perquisite emotion types), and asked to fill a questioner on how they perceived the emotion in the images. The questioner consists of 16 DES emotions [21-23] scales from 0 to 100 correspond to nonexistence to fully identified.

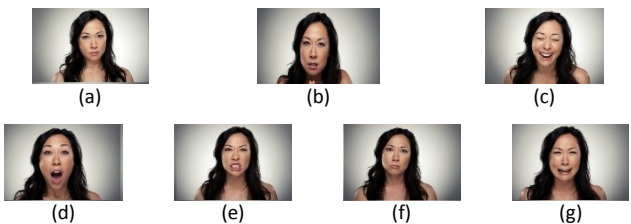

Fig. 1. The seven expresive facials images used in the experiments. All participants are asked to identify the emotions expressed by the person in the images.

As participants' emotion baseline, all participants are required to fill their own emotion prior the experiment (i.e. the participants' own initial emotions, and will be called *IE* throughout this article).

### B. Statistical methods

All data collected from the survey will be processed through a series of statistical calculation to be analyzed. The analysis focuses on answering the three hypotheses discussed shortly in this section. Yet, it is important to note that the experiment does not assume the type of the emotions showed in the images; hence the participants' recognition accuracies do not need to be evaluated. All statistical calculations are implemented using R version 3.2.3.

The statistic calculations are conducted accordingly to answer the three RQs:

*1) Do participants able to take the perspective of other?*

Prior using emotionally expressive facial images in research, a basic question should be asked whether the participants are able to take the perspective of other. Though people are capable to make a clear distinction between their own and other perspective, there is indication that this capability may not be the same for everyone (cf. [16-18]).

The relations of the participants' emotion and the one they perceived from the images will be evaluated using student t-test, where a non-significant test should demonstrate similar emotions between the two and therefore indicate participants' inability to take the perspective of the person in the images. And so the first hypothesis is ($H_{0(1)}$) that the participants are able to take the perspective of other.

*2) Do participants collectively appoint to the same emotions shown in the images?*

Most HCI studies that use facial expressive images assume that the all expression correspond uniquely to one targeted emotion label. When should the assumption failed, the users (experimenter/participants) could wrongly perceive other emotion than the targeted emotion (cf. [16-19]).

The recognition rates are measured by calculating student t-test on all emotions. Significantly strong emotion signifies a collective perceived emotion (the DES scores which will be called the *RE* throughout the article) concession between participants. Thus the second hypothesis is ($H_{0(2)}$) that the participants do collectively recognize the images of having the same emotions.

*3) Do men and women appoint to the same emotions depicted by the images?*

The next step of investigation is to look how men and women perceived the targeted emotion. Again, most HCI studies assume that men and women similarly perceive the emotions whereas other studies in behavior science suggest that women have higher sensitivity on affective interaction [19].

Another student t-test measurement will be applied on the *RE* score between men and women. The test should reveal whether men and women significantly differ in perceiving emotion. Therefore, the last hypothesis is ($H_{0(3)}$) that there is no difference between men and women on perceiving emotions showed in the images.

## IV. RESULTS

The results of the experiment will be presented in this section in the following order: (a) the description of participants' *IE*, (b) the differences between the *IE* and *RE* to evaluate whether the participants *IE* are affecting their judgment of their *RE* (answer to RQ1), (c) the (dis)similarity of participants *RE* (answer to RQ2), and (d) the (dis)similarity of the *RE* between men and women (answer to RQ3).

### A. Participants initial emotion (IE)

Prior to hypothesis tests, the *IE* of all participants are tested using student t-test to evaluate their own feeling before examining the images in Figure 1. This is an important step for evaluating whether or not the *RE* is a reflection of the *IE*. In the case of *RE* is a reflection of *IE* then the *RE* will not significantly differ from the *IE*, which automatically answer RQ1 and suggest a link between the *IE* and the answers of RQ2 and RQ3.

In general, the participants do not show particular elevated emotions prior the experiments (Fig. 2a) with $\mu = 23.8, \sigma = 28.6$, except for two emotions where they slightly elevated (though not significant): calm ($\mu = 59.8, \sigma = 25.2$) and happy ($\mu = 57.2, \sigma = 23.6$) which suggest a positive attitude toward the experiment. The absence of significant intensity in all of the 16 emotions implies participants' neutral emotional states [24] before the experiment begin.

The *IE* felt by the participants are varied demonstrated by their standard deviation ($5.5 \leq \sigma \leq 37.2$). An interesting pattern is showed when comparing the mean and the standard deviation of the initial 16 emotions (Fig. 2b). The participants show a decreasing variation towards the two extreme of emotion's intensity. The participants tend to answer common scores when the *IE* are in the two extreme intensities, and most varied when the intensity is in between. Although this experiment is not intended to check the consistency of emotions how people perceive emotions, the ascending and descending pattern of the emotions' intensity could indicate a binary perception of emotions (i.e. emotions are easier to perceive when in the two extremes). Similar pattern also appears where the participants show a similar emotion intensities of the *RE* when the intensities are in the two extremes (discussed in Section III.C).

### B. (RQ1) Do participants able to take the perspective of other?

The neutrality of the participants' *IE* should suggest that any participants' significantly elevated *RE* should not be caused by their *IE*. As a result, any significant increment of the *RE* from the *IE* (i.e. $RE \gg IE$) should proof that *IE* is not affecting *RE*, and thus the participants are able to perceive different emotions than what they are feeling. The $RE \gg IE$ condition for each image is determined by fulfilling:

$$RE_i \gg IE_i \leftarrow (d_i > 0) \wedge (p_i \leq 0.05) \quad (1)$$

where $d_i$ and $p_i$ are the dissimilarity, and the $p$-value between the *RE* and *IE* respectively for each $i$ emotions.

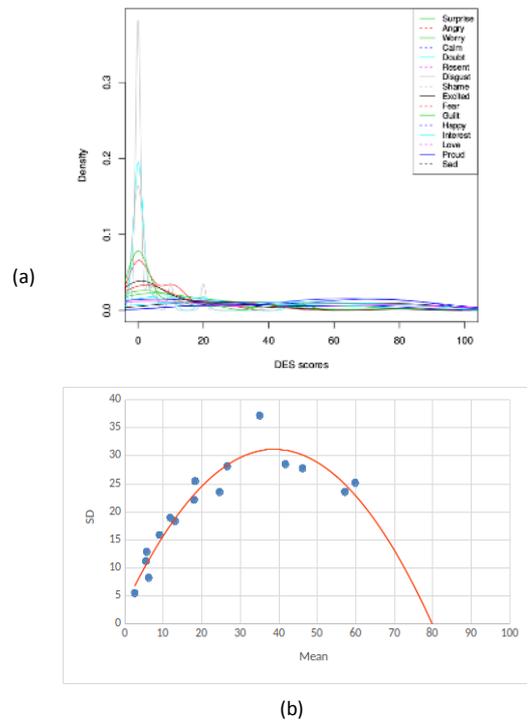

Fig. 2. (a) Density plot of participants' *IE* states measured by 16 emotions. Participants showing a relatively netural emotions indicated by the near-zero DES values on all emotions. (b) The average and the standard deviation of the initial 16 measured emotions. The red line in (b) is the second order polynomial regression's trendline ($R^2 = 0.91$).

TABLE I.   MEAN DIFFERENCES BETWEEN THE *IE* AND *RE* FROM FIGURE 1. THE NUMBERS ARE THE DIFFERENCES BETWEEN THE *RE* INTENSITY AND *IE*'S. POSITIVE VALUE INDICATES $RE > IE$. YELLOW BACKGROUND INDICATES SIGNIFICANT VARIANCE DIFFERENCES BETWEEN *RE* AND *IE*. **BOLDFACE** SIGNIFY $RE \gg IE$.

| Emotions | Images in Figure 1 | | | | | | |
|---|---|---|---|---|---|---|---|
| | *(a)* | *(b)* | *(c)* | *(d)* | *(e)* | *(f)* | *(g)* |
| **Surprise** | -5.9 | 6.7 | 2.9 | **70.6** | 12.1 | **21.9** | 5.3 |
| **Angry** | **38.2** | **20.5** | -5.5 | 7.6 | **77.7** | **51.2** | **10.4** |
| **Worry** | **14.9** | **24.5** | -13.8 | -3.3 | 5.6 | **18.6** | **18.6** |
| **Calm** | -30.8 | -51.1 | -27.7 | -50.2 | -58.2 | -53.1 | -58.3 |
| **Doubt** | 2.1 | 9.1 | -19.9 | -17.1 | -7.5 | 3.8 | -1.6 |
| **Resent** | **38.6** | **27.7** | -3.5 | 3.0 | **50.2** | **51.4** | **13.0** |
| **Disgust** | **10.5** | **42.3** | -1.6 | 7.7 | **49.5** | **12.1** | **12.2** |
| **Shame** | 2.8 | 5.7 | 1.1 | 3.4 | -0.9 | 5.7 | **19.0** |
| **Excited** | -39.1 | -34.9 | 12.8 | 12.6 | -23.7 | -40.9 | -41.8 |
| **Fear** | 7.4 | **25.9** | -10.2 | -1.3 | -0.1 | 4.2 | **28.6** |
| **Guilt** | 9.9 | **26.2** | -6.9 | -3.1 | -1.3 | 4.2 | **29.9** |
| **Happy** | -48.5 | -51.9 | **34.1** | -17.5 | -55.8 | -54.6 | -56.8 |
| **Interest** | -34.5 | -31.4 | 10.1 | 1.3 | -37.4 | -32.0 | -35.7 |
| **Love** | -30.6 | -31.2 | **19.9** | -18.6 | -34.1 | -32.2 | -27.6 |
| **Proud** | -18.7 | -22.5 | **20.5** | -1.5 | -23.3 | -23.4 | -23.6 |
| **Sad** | 12.8 | 6.8 | -11.1 | -8.2 | 2.3 | 8.8 | **75.2** |

The experiment shows that all expressive facial images have at least one $RE$ that fulfills the condition in Equation 1 (**boldface** in Table 1). This suggest that there are at least one emotion that are not related to what the participants are feeling (i.e. the participants are able to distinguish between their own feeling and the feeling expressed in the images), and therefore implies the rejection of $H_{0(1)}$. It is concluded that the participants are able to take the perspective of others regardless of their own feeling at the time.

*C. (RQ2) Do participants collectively appoint to the same emotions shown in the images?*

The present of emotions are indicated by the appearance of dominance emotions (e.g. surprise emotion is indicated by a very high surprise intensity relative to the other emotions) (cf. [24]). These indications are expressed by high intensity and distinct emotions (i.e. high and statistically different DES values). By this definition, the RQ2 can be answer by evaluating the $RE$ for each image by pointing to high positive and significant DES values:

$$RE_i \gg RE_j \leftarrow \left(\max\left(e_{(i,j)}\right)\right) \wedge \left(p_{(i,j)} \leq 0.05\right) \quad (2)$$

where $i$ and $j$ are the emotions with $i \neq j$, and $e$ and $p$ correspond to the dissimilarity and the $p$-value of the $RE$. The condition where the participants collectively perceive comparable emotions occurs when the condition in Equation 2 is met.

Figure 3 shows the relation between the mean and the standard deviation of the 16 $RE$ in all seven images. By visual inspection, it is clear that only a few emotions are distinctly perceived by the participants (i.e. demonstrated by the cluttered emotions on the small-medium side of the DES values).

Those few emotions appear only in four images (c, d, e, and g) which are shown in detail in Table 2. The table presents the intensity and the number of significantly different $RE$ ranging between zero to 15, correspond to the minimum and the maximum number of $RE$ combinations (i.e. zeros implies no differences between the $RE_i$ and the $RE_j$, while 15 implies a significant $RE_i$ amongst the $RE_j$). The **boldface** values signify the condition in Equation 2.

Based on the result described in Table 2, it concludes the partial rejection of $H_{0(2)}$. The rejections occur in four images (c, d, e, and g) out of seven images. This outcome suggests that not all expressive facial expressions are constantly perceived universally and some emotional ambiguity should be expected when introducing facial expressions images in research. Participants seem to have more alignment in regards to the Ekman's six basic emotions [25-26] than other emotions, which support the Ekman's idea of the six basic emotions as building blocks to other emotions.

An unexpected finding appears when comparing the mean and the standard deviation of the $RE$ (Figure 3). They appear to follow a trend line where the standard deviation is relatively smaller on the small and large values of mean. This trend suggests that participants are more agreeable on the two extreme of emotions, forming a binary decision (i.e. whether or not the image is showing an emotion). Similar pattern also occurs in the $IE$ (Figure 2). Although this experiment is not intended to check nor to explore the consistency of perceived emotions (self and other emotions), this finding may not be arbitrary and should be followed in the future studies.

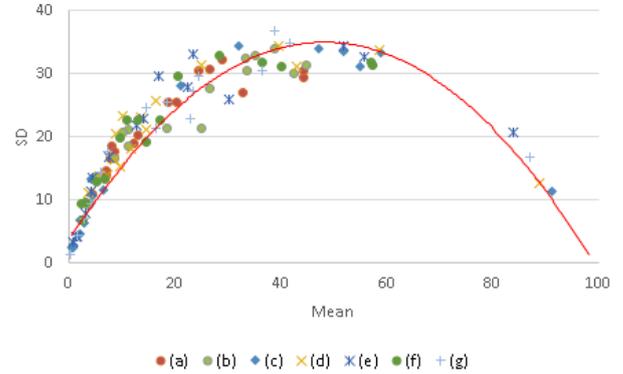

Fig. 3. The average and the standard deviation of the $RE$. The red line in is the second order polynomial regression's trendline ($R^2 = 0.97$). Notice that the relation between the average and the standard deviation of the $RE$ is similar with the one in the $IE$ (Figure 2b).

TABLE II. THE PERCEIVED EMOTIONS ($RE$) OF THE IMAGES IN FIGURE 1.

| Emotions | Images in Figure 1 | | | | | | |
|---|---|---|---|---|---|---|---|
| | (a) | (b) | (c) | (d) | (e) | (f) | (g) |
| Surprise | 12.4 (5) | 25.0 (9) | 21.2 (14) | **88.9 (15)** | 30.3 (12) | 40.2 (13) | 23.5 (8) |
| Angry | 44.4 (12) | 26.6 (9) | 0.7 (8) | 13.7 (5) | **83.9 (15)** | 57.3 (14) | 16.5 (9) |
| Worry | 32.9 (9) | 42.5 (10) | 4.2 (7) | 14.7 (5) | 23.6 (10) | 36.6 (13) | 36.6 (10) |
| Calm | 29.0 (8) | 8.7 (9) | 32.1 (13) | 9.6 (5) | 1.6 (9) | 6.8 (7) | 1.5 (10) |
| Doubt | 26.7 (10) | 33.7 (8) | 4.6 (7) | 7.5 (5) | 17.1 (9) | 28.4 (10) | 23.0 (9) |
| Resent | 44.3 (12) | 33.4 (8) | 2.2 (7) | 8.7 (5) | 55.9 (14) | 57.1 (14) | 18.7 (9) |
| Disgust | 13.1 (6) | 44.8 (10) | 1.0 (8) | 10.2 (5) | 52.0 (14) | 14.7 (5) | 14.7 (8) |
| Shame | 8.2 (7) | 11.2 (8) | 6.5 (10) | 8.9 (5) | 4.5 (7) | 11.1 (5) | 24.5 (8) |
| Excited | 7.0 (8) | 11.3 (8) | 59.0 (12) | 58.8 (14) | 22.5 (10) | 5.3 (8) | 4.3 (11) |
| Fear | 20.4 (8) | 38.9 (8) | 2.9 (7) | 11.7 (6) | 12.9 (8) | 17.2 (9) | 41.6 (13) |
| Guilt | 18.9 (6) | 35.2 (8) | 2.1 (7) | 5.9 (7) | 7.7 (7) | 13.2 (8) | 38.9 (10) |
| Happy | 8.6 (7) | 5.3 (9) | **91.2 (15)** | 39.7 (13) | 1.3 (9) | 2.5 (9) | 0.3 (14) |
| Interest | 7.2 (8) | 10.3 (8) | 51.8 (12) | 43.0 (13) | 4.3 (8) | 9.7 (5) | 6.0 (10) |
| Love | 4.4 (9) | 3.8 (9) | 54.9 (12) | 16.4 (5) | 0.9 (10) | 2.8 (9) | 7.5 (8) |
| Proud | 7.9 (7) | 4.1 (9) | 47.1 (11) | 25.1 (11) | 3.3 (9) | 3.1 (9) | 3.0 (11) |
| Sad | 24.6 (8) | 18.6 (10) | 0.7 (8) | 3.6 (9) | 14.1 (9) | 20.6 (9) | **87.0 (15)** |

*D. (RQ3) Do men and women appoint to the same emotions depicted by the images?*

To answer RQ3, the data is divided based on participants' gender (male and female) and apply student t-test on each of the DES emotion types between the two genders. The experiments found that $p \geq 0.05$ in all tests which conclude an acceptance of $H_{0(3)}$ (i.e. both men and women concordantly perceive the same type of emotion illustrated in the images).

## V. DISCUSSION

The experiments introduce the answers for the RQs and find one secondary result. In the light of the first RQ, the participants are able to perceived facial expression regardless of their own emotions. The ability of the participants to take the perspective of others' feeling regardless of their own confirms the theorem suggested by [14] and [15]. This result also verifies the usefulness of facial images in emotion related research use. However, researchers should aware that not all participants will perceive the same emotions for all facial expressions.

For the second RQ, people appear to be able to take the perspective of others better on certain expressions but not the others (e.g. the expressions in images (c-e, g) in Figure 1). This behavior appears to agree with the theory of Ekman [25-26] on six basic emotions which are strongly related to facial expressions (i.e. FACS [27]). Based on this result (i.e. the fact that the four successfully induced emotions are affiliated with Ekman basic emotions), it appears that it is relatively safe to use facial expression images related to the six basic emotions for research.

As for the third RQ, there is no evidence that men and women have different ability in taking the perspective of others, contradicting [19].

A secondary result is shown by the grouping pattern between the mean and the standard deviation on the perceived emotions. There is more agreement between participants when the perceived emotions are in the binary intensity (very low or very high) than when they perceived the emotions of having intermediate intensity. This would suggest the importance of using intense expressions in emotion related research. Nevertheless, further studies are still needed to confirm the occurrence of this behavior.

Two remarks should be noted regarding the findings. First, our participants consist of higher education students who have relatively high educational background. Since there is a chance that education might affect the perception of emotions, further validation is needed to confirm the finding on more varies participants. Second, the findings are based on the requirement function (e.g. Equation 1 and 2) where the experiment set the probability significant value threshold as 0.05 ($p \leq 0.05$). A less sensitive threshold (e.g. $p \leq \{0.1, \ldots, 0.3\}$) might suggest different results, yet with the cost of oversensitive emotion detection.

## VI. CONCLUSION

The experiments conclude the answer of the three RQs. In general, the experiments find that people are able to take the perspective of others on the six basic emotions. This ability appears to work regardless of their gender and their own emotion at the time, which therefore conclude that it is relatively safe to use facial images to elicit emotions as long the targeted emotions are within the scope of Ekman's six basic emotions.

Future study should takes the experiment further than the scope of the basic emotions, and explore the binary perception of emotions behavior introduced in Section IV.A and IV.C.

## ACKNOWLEDGMENT

The data collected in this study is part of the theses by Aurelia Rianto (Computer Science Dept.) and Febiardi (Information System Dept.). This study is under grant number 265/LPPM/UMN/III/2017 from Universitas Multimedia Nusantara. The author would like to thank Friska Natalia Ferdinand, PhD for her constructive feedback of the manuscript.

## REFERENCES

[1] C. Busso, Z. Deng, S. Yildirim, M. Bulut, C. M. Lee, A. Kazemzadeh, S. Lee, U. Neumann, and S. Narayanan, "Analysis of emotion recognition using facial expressions, speech and multimodal information," Proceedings of the 6th international conference on Multimodal interfaces, ACM, pp. 205-211, October 2004.

[2] R. Mattheij, M. Nilsenova, and E. Postma, "Vocal and facial imitation of humans interacting with virtual agents," Humaine Association Conference on Affective Computing and Intelligent Interaction (ACII),pp. 815-820, September 2013.

[3] R. Mattheij, M. Postma-Nilsenová, and E. Postma, "Mirror mirror on the wall," Journal of Ambient Intelligence and Smart Environments, 7(2), pp. 121-132, 2015.

[4] N. B. Wunarso and Y. E. Soelistio, "Towards Indonesian speech-emotion automatic recognition (I-SpEAR)," 2017 4th International Conference on New Media Studies (CONMEDIA), Yogyakarta, 2017, pp. 98-101. doi: 10.1109/CONMEDIA.2017.8266038

[5] M. S. Bartlett, G. C. Littlewort, M. G. Frank, and K. Lee, K., "Automatic decoding of facial movements reveals deceptive pain expressions," Current Biology, 24(7), pp. 738-743, 2014.

[6] C, Lamm, E. C. Porges, J. T. Cacioppo, andJ. Decety, "Perspective taking is associated with specific facial responses during empathy for pain,"Brain research, 1227, pp. 153-161, 2008.

[7] Y. B. Sun, Y. Z. Wang, J. Y. Wang, and F. Luo, "Emotional mimicry signals pain empathy as evidenced by facial electromyography," Scientific reports, 5, 16988, 2015.

[8] W. Sato, T. Kochiyama, and S. Uono, "Spatiotemporal neural network dynamics for the processing of dynamic facial expressions,"Scientific reports, 5, 12432, 2015.

[9] H. Aviezer, Y. Trope, and A. Todorov, "Body cues, not facial expressions, discriminate between intense positive and negative emotions," Science, 338(6111), pp. 1225-1229, 2012.

[10] S. Wu, and B. Keysar, "The effect of culture on perspective taking," Psychological science, 18(7), pp. 600-606, 2007.

[11] A. M. Kring, and A. H. Gordon, "Sex differences in emotion: expression, experience, and physiology," Journal of personality and social psychology, 74(3), pp. 686-703, 1998.

[12] D. McDuff, E. Kodra, R. el Kaliouby, and M. LaFrance, "A large-scale analysis of sex differences in facial expressions," PloS one, 12(4), e0173942, 2017.

[13] W. J. Chopik, E. O'Brien, and S. H. Konrath, "Differences in empathic concern and perspective taking across 63 countries,"Journal of Cross-Cultural Psychology, 48(1), pp. 23-38, 2017

[14] B. Underwood, andB. Moore, "Perspective-taking and altruism.,"Psychological bulletin, 91(1), 1982.

[15] C. D. Batson, K. Sager, E. Garst, M. Kang, K. Rubchinsky,and K. Dawson,"Is empathy-induced helping due to self–other merging?,"Journal of personality and social psychology, 73(3), 1997.

[16] C. D. Batson, J. H. Eklund, V. L. Chermok, J. L. Hoyt, and B. G. Ortiz, "An additional antecedent of empathic concern: valuing the welfare of the person in need," Journal of personality and social psychology, 93(1), pp. 65-74, 2007.

[17] S. Wu, andB. Keysar,"The effect of culture on perspective taking.,"Psychological science, 18(7), pp. 600-606, 2007.

[18] R. E. Jack, O. G. Garrod, H. Yu, R. Caldara, andP. G. Schyns,"Facial expressions of emotion are not culturally universal,"Proceedings of the National Academy of Sciences, 109(19), pp. 7241-7244, 2012.

[19] A.M. Kring, andA. H. Gordon,"Sex differences in emotion: expression, experience, and physiology,"Journal of personality and social psychology, 74(3), pp. 686-703, 1998

[20] mikelarremore.com, 'Faceboards', 2017. [Online]. Available: http://mikelarremore.com/faceboards. [Accessed: 10- Mei- 2017].

[21] C. E. Izard, F. E. Dougherty, B. M. Bloxom, and N. E. Kotsch, "The Differential Emotions Scale: a Method of Measuring the Meaning of Subjective Experience of Discrete Emotions," Nashville, TN: Vanderbilt University, Department of Psychology, 1974.

[22] G. J. McHugo, C. A. Smith, and J. T. Lanzetta, "The Structure of Self-reports of Emotional Responses to Film Segments," Motivation & Emotion, 6(4), pp. 365-385, 1982.

[23] P. Philippot, "Inducing and Assessing Differentiated Emotional Feeling States in The Laboratory," Cognition and Emotion, 7, pp. 171-193, 1993.

[24] X. Qiao, H. Li, J. Xiang, andH. Deng,"The study of images emotion based on fMRI," International Conference on Web Information Systems and Mining, Springer, Berlin, Heidelberg, pp. 66-72, 2011.

[25] P. Ekman," An argument for basic emotions," Cognition & emotion, 6(3-4), pp. 169-200, 1992.

[26] P. Ekman, "Are there basic emotions?," 1992.

[27] P. Ekman and W. Friesen, "The Facial Action Coding System: A Technique For The Measurement of Facial Movement," Consulting Psychologists Press, Inc., San Francisco, CA, 1978.

[28] M. Wegrzyn, M. Vogt, B. Kireclioglu, J. Schneider, and J. Kissler, "Mapping the emotional face. How individual face parts contribute to successful emotion recognition," PloS one, 12(5), e0177239, 2017.

[29] S. Sullivan, A. Campbell, S. B. Hutton, and T. Ruffman, "What's good for the goose is not good for the gander: Age and gender differences in scanning emotion faces," The Journals of Gerontology: Series B, 72(3), 441-447, 2017.